\documentclass[10pt,conference]{IEEEtran}
\usepackage{amssymb,amsmath} 
\usepackage{cite}
\usepackage{paralist}

\newcommand{\C}{\mathbf{C}}
\newcommand{\F}{\mathbf{F}}
\newcommand{\NN}{\mathbf{N}}
\newcommand{\N}{\mathcal{N}}

\newcommand{\wt}{{\rm {wt}}}

\newcommand{\hdual}{{\bot_h}}
\newcommand{\scal}[2]{\langle #1\,|\,#2\rangle}
\newcommand{\zero}{\mathbf{0}}
\newcommand{\one}{\mathbf{1}}
\newcommand{\RM}{{\mathcal{R}}}
\newcommand{\floor}[1]{\left\lfloor {#1}\right\rfloor}
\newcommand{\ceil}[1]{\left\lceil {#1}\right\rceil}
\newcommand{\ket}[1]{|#1\rangle}
\newcommand{\nix}[1]{}

\DeclareMathOperator{\tr}{tr} 
\DeclareMathOperator{\dirlimit}{{\displaystyle\lim_{\longrightarrow}}}

\newtheorem{theorem}{Theorem}
\newtheorem{lemma}[theorem]{Lemma}
\newtheorem{definition}[theorem]{Definition}

\newtheorem{corollary}[theorem]{Corollary}

\begin{document}
\title{Quantum Convolutional Codes Derived from Reed-Solomon and Reed-Muller Codes}
\author{
\authorblockN{Salah A. Aly, Andreas Klappenecker, Pradeep Kiran Sarvepalli}
\authorblockA{Department  of Computer Science, Texas A\&M University\\
College Station, TX 77843-3112, USA \\
Email: \{salah, klappi, pradeep\}@cs.tamu.edu}} 

\maketitle

\begin{abstract}
Convolutional stabilizer codes promise to make quantum communication
more reliable with attractive online encoding and decoding
algorithms. This paper introduces a new approach to convolutional
stabilizer codes based on direct limit constructions. Two families of
quantum convolutional codes are derived from generalized
Reed-Solomon codes and from Reed-Muller codes. A Singleton bound for
pure convolutional stabilizer codes is given.
\end{abstract}

\section{Introduction}
A key obstacle to the communication of quantum information is
decoherence, the spontaneous interaction of the environment with the
information-carrying quantum system. The protection of quantum
information with quantum error-correcting codes to reduce or perhaps
nearly eliminate the impact of decoherence has led to a highly
developed theory of quantum error-correcting block codes. Somewhat
surprisingly, quantum convolutional codes have received less
attention.

Ollivier and Tillich developed the stabilizer framework for quantum
convolutional codes, and addressed encoding and decoding aspects of
such codes~\cite{ollivier03,ollivier04}. Almedia and Palazzo
constructed a concatenated convolutional code of rate $1/4$ with
memory $m=3$~\cite{almeida04}. Forney and Guha constructed quantum
convolutional codes with rate $1/3$ \cite{forney05a}. Also, in a joint
work with Grassl, they derived rate $(n-2)/n$ convolutional stabilizer
codes~\cite{forney05b}. Grassl and R{\"{o}}tteler constructed quantum
convolutional codes from product codes~\cite{grassl05}, and they 
gave a general algorithm to obtain
non-catastrophic encoders~\cite{grassl06}.

In this paper, we give a new approach to quantum convolutional codes
based on a direct limit construction, generalize some of the
previously known results, and construct two families of quantum
convolutional codes based on classical generalized Reed-Solomon and
Reed-Muller codes.


\section{Background}\label{SEC:background}
In this section, we give some background concerning classical
convolutional codes, following \cite[Chapter 14]{huffman03} and
\cite{lally06}.

Let $\F_q$ denote a finite field with $q$ elements.  An
$(n,k,\delta)_q$ \textit{convolutional code} $C$ is a submodule of
$\F_q[D]^n$ generated by a right-invertible matrix $G(D)=(g_{ij})\in
\F_q[D]^{k\times n}$,
\begin{eqnarray}\label{eq:cc-def}
C =\{  \textbf{u}(D) G(D) \mid \mathbf{u}(D) \in \F_q[D]^k\},
\end{eqnarray}
such that $\sum_{i=1}^k \nu_i = \max\{ \deg \gamma\,|\, \gamma \text{
is a $k$-minor of $G(D)$}\}$ $=:\delta$, where $\nu_i = \max_{1\leq
j\leq n} \{\deg g_{ij} \}$. We say $\delta$ is the \textit{degree} of $C$.
The \textit{memory} $\mu$ of $G(D)$ is
defined as $\mu=\max_{1\le i\le k} \nu_i$.  The \textit{weight}
$\wt(v(D))$ of a polynomial $v(D)$ in $\F_q[D]$ is defined as the
number of nonzero coefficients of $v(D)$, and the \textit{weight} of
an element $\mathbf{u}(D)\in \F_q[D]^n$ is defined as
$\wt(\mathbf{u}(D))=\sum_{i=1}^n \wt(u_i(D))$.  The \textit{free
distance} $d_f$ of $C$ is defined as $d_f =\wt(C)=\min \{ \wt (u)\mid u \in
C, u\neq 0 \}.$ We say that an $(n,k,\delta)_q$ convolutional code
with memory $\mu$ and free distance $d_f$ is an $(n,k,\delta;\mu,d_f)_q$
convolutional code.

Let $\NN$ denote the set of nonnegative integers. Let $\Gamma_q= \{
v\colon \NN\rightarrow \F_q\,|\, \text{ all but finitely many
coefficients of $v$ are 0}\}$. We define a vector space isomorphism
$\sigma\colon \F_q[D]^n\rightarrow \Gamma_q$ that maps an element
$\mathbf{u}(D)$ in $\F_q[D]^n$ to the coefficient sequence of the polynomial
$\sum_{i=0}^{n-1} D^i u_i(D^n)$, that is, an element in $\F_q[D]^n$ is
mapped to its interleaved coefficient sequence. Frequently, we will
refer to the image $\sigma(C)$ of a convolutional code
(\ref{eq:cc-def}) again as $C$, as it will be clear from the context
whether we discuss the sequence or polynomial form of the code. Let
$G(D)= G_0 + G_1 D +\cdots + G_\mu D^\mu$, where $G_i\in \F_q^{k\times n}$
for $0\le i\le \mu$. We can associate to the generator matrix $G(D)$ its
semi-infinite coefficient matrix
\begin{eqnarray}\label{eq:Gmatrix}
G= \begin{pmatrix}
       G_0 & G_1 & \cdots & G_\mu & & \\
        & G_0 & G_1 & \cdots & G_\mu &  \\
        &  & \ddots  & \ddots &  & \ddots \\
     \end{pmatrix}.
\end{eqnarray}
If $G(D)$ is the generator matrix of a convolutional code $C$, then
one easily checks that $\sigma(C)=\Gamma_q G$.

In the literature, convolutional codes are often defined in the form
$\{ p(D)G'(D)\,|\, p(D)\in \F_{q}(D)^k\}$, where $G'(D)$ is a matrix of
full rank in $\F_q^{k\times n}[D]$. In this case, one can obtain a
generator matrix $G(D)$ in our sense by multiplying $G'(D)$ from the
left with a suitable invertible  matrix $U(D)$ in $\F_q^{k\times k}(D)$, see~\cite{huffman03}.

We define the \textit{Euclidean inner product} of two sequences $u$ and $v$ in
$\Gamma_q$ by
$ \scal{u}{v} = \sum_{i\in \NN} u_iv_i$, and the Euclidean dual
of a convolutional code $C\subseteq \Gamma_q$ by 
$C^\perp=\{ u\in \Gamma_q\,|\, \scal{u}{v}=0 \text{ for all } v\in C\}$.
A convolutional code $C$ is called self-orthogonal if and only if
$C\subseteq C^\perp$. It is easy to see that a convolutional code $C$ is
self-orthogonal if and only if $GG^T=0$.

Consider the finite
field $\F_{q^2}$. The \textit{Hermitian inner product} of two
sequences $u$ and $v$ in $\Gamma_{q^2}$ is defined as $ \scal{u}{v}_h
= \sum_{i\in \NN} u_i\, v_i^q.$ We have $C^\hdual = \{ u\in
\Gamma_{q^2}\,|\, \scal{u}{v}_h=0 \text{ for all } v\in C\}$.  As
before, $C\subseteq C^\hdual$ if and only if $GG^\dagger=0$, where the
Hermitian transpose $\dagger$ is defined as $(a_{ij})^\dagger =
(a_{ji}^q)$.


\section{Quantum Convolutional Codes}\label{sec:QCCparameters}
The state space of a $q$-ary quantum digit is given by the complex
vector space $\C^q$. Let $\mbox{$\{ \ket{x}\,|\, x\in \F_q\}$}$ denote
a fixed orthonormal basis of $\C^q$, called the computational
basis. For $a,b\in \F_q$, we define the unitary operators
$$ X(a)\ket{x}=\ket{x+a}\;\; \text{and}\;\; Z(b)\ket{x}=\exp(2\pi i
\tr(bx)/p)\ket{x},$$ where the addition is in $\F_q$, $p$ is the
characteristic of $\F_q$, and $\tr(x)=x^p+x^{p^2}+\cdots+x^q$ is the
absolute trace from $\F_q$ to $\F_p$. The set
$\mathcal{E}=\{X(a),Z(b)\,|\, a,b\in\F_q\}$ is a basis of the algebra
of $q\times q$ matrices, called the \textit{error basis}.

A quantum convolutional code  encodes a stream of quantum digits. One
does not know in advance how many qudits {\em i.e.}, quantum digits will be sent, so the idea is
to impose structure on the code that simplifies online encoding and
decoding. Let $n$, $m$ be positive integers. We will process $n+m$
qudits at a time, $m$ qudits will overlap from one step to the next,
and $n$ qudits will be output.

For each $t$ in $\NN$, we define the Pauli group
$P_t=\langle M | M\in \mathcal{E}^{\otimes (t+1)n+m}\rangle$ as the group
generated by the \mbox{$(t+1)n+m$}-fold tensor product of the error
basis~$\mathcal{E}$. Let $I=X(0)$ be the $q\times q$ identity matrix.  
For $i,j\in \NN$ and $i\le j$, we define the inclusion
homomorphism $\iota_{ij}\colon P_i\rightarrow P_j$ by
$\iota_{ij}(M)=M\otimes I^{\otimes n(j-i)}$. We have
$\iota_{ii}(M)=M$ and $\iota_{ik}=\iota_{jk}\circ \iota_{ij}$ for
$i\le j\le k$. Therefore, there exists a group 
$$ P_\infty = \dirlimit (P_i,\iota_{ij}),$$ called the direct limit of
the groups $P_i$ over the totally ordered set $(\NN,\le)$.  For each
nonnegative integer $i$, there exists a homomorphism $\iota_i\colon
P_i\rightarrow P_\infty$ given by $\iota_i(M_i)=M_i\otimes I^{\otimes
\infty}$ for $M_i\in P_i$, and $\iota_i =\iota_j\circ \iota_{ij}$
holds for all $i\le j$. We have $P_\infty = \bigcup_{i=0}^\infty
\iota_i(P_i)$; put differently, $P_\infty$ consists of all infinite
tensor products of matrices in $\langle M\,|\,M\in \mathcal{E}\rangle$
such that all but finitely many tensor components are equal to $I$.
The direct limit structure that we introduce here provides the proper
conceptual framework for the definition of convolutional stabilizer
codes; see~\cite{ribes00} for background on direct limits. 

We will define the stabilizer of the quantum convolutional code also
through a direct limit. Let $S_0$ be an abelian subgroup of $P_0$. For
positive integers $t$, we recursively define a subgroup $S_t$ of $P_t$ by
$S_t=\langle N\otimes I^{\otimes n}, I^{\otimes tn}\otimes M\,|\,
N\in S_{t-1}, M\in S_0\rangle.$
Let $Z_t$ denote the center of the group $P_t$. We will assume that 
\begin{compactenum}[\bf S1)]
\item $I^{\otimes
tn} \otimes M$ and $N\otimes I^{\otimes tn}$ commute for all $N,M\in
S_0$ and all positive integers $t$. 
\item $S_tZ_t/Z_t$ is an
$(t+1)(n-k)$-dimensional vector space over $\F_q$. 
\item $S_t\cap Z_t$ contains only the identity
matrix.
\end{compactenum} Assumption \textbf{S1} ensures that $S_t$ is
an \textit{abelian} subgroup of~$P_t$, \textbf{S2} implies that $S_t$
is generated by $t+1$ shifted versions of $n-k$ generators of $S_0$
and all these $(t+1)(n-k)$ generators are independent, and \textbf{S3}
ensures that the stabilizer (or $+1$ eigenspace) of $S_t$ is
nontrivial as long as $k<n$.

The abelian subgroups $S_t$ of $P_t$ define an abelian group   
$$S=\dirlimit (S_i,\iota_{ij})= \langle \iota_t(I^{\otimes tn}\otimes
M)\,|\, t\ge 0, M\in S_0\rangle$$
generated by shifted versions of elements in $S_0$. 
\begin{definition}
Suppose that an abelian subgroup $S_0$ of $P_0$ is chosen such that 
\textbf{S1}, \textbf{S2}, and \textbf{S3} are satisfied. 
Then the $+1$-eigenspace of
$S=\displaystyle\lim_{\longrightarrow} (S_i,\iota_{ij})$ in
$\bigotimes_{i=0}^\infty \C^q$ defines a convolutional
stabilizer code with parameters $[(n,k,m)]_q$.
\end{definition}
\smallskip

In practice, one works with a stabilizer $S_t$ for some large (but
previously unknown) $t$, rather than with $S$ itself. We notice that
the rate $k/n$ of the quantum convolutional stabilizer code defined by
$S$ is approached by the rate of the stabilizer block code $S_t$ for
large $t$. Indeed, $S_t$ defines a stabilizer code with
parameters $[[(t+1)n+m,(t+1)k+m]]_q$; therefore, the rates of these
stabilizer block codes approach
$$ 
\lim_{t\rightarrow \infty} \frac{(t+1)k+m}{(t+1)n+m} = 
\lim_{t\rightarrow \infty} \frac{k+m/(t+1)}{n+m/(t+1)} = \frac{k}{n}.$$  

We say that an error $E$ in $P_\infty$ is \textit{detectable} by a
convolutional stabilizer code with stabilizer $S$ if and only if a
scalar multiple of $E$ is contained in $S$ or if $E$ does not commute
with some element in $S$. The \textit{weight} $\wt$ of an element in
$P_\infty$ is defined as its number of non-identity tensor components.
A quantum convolutional stabilizer code is said to have \textit{free
distance} $d_f$ if and only if it can detect all errors of weight less
than $d_f$, but cannot detect some error of weight $d_f$.  Denote by
$Z(P_\infty)$ the center of $P_\infty$ and by $C_{P_\infty}(S)$ the
centralizer of $S$ in $P_\infty$.  Then the free distance is given by
$d_f = \min \{\wt(e)\mid e\in C_{P_\infty}(S)\setminus
Z(P_\infty)S\}$.

Let $(\beta,\beta^q)$ denote a normal basis of $\F_{q^2}/\F_q$. Define
a map $\tau\colon P_\infty\rightarrow \Gamma_{q^2}$ by 
$\tau(\omega^cX(a_0)Z(b_0)\otimes X(a_1)Z(b_1)\otimes \cdots)= 
(\beta a_0+\beta^q b_0,\beta a_1+\beta^q b_1,\dots)$.
For sequences $v$ and $w$ in
$\Gamma_{q^2}$, we define a trace-alternating form
$$ 
\scal{v}{w}_a = \tr_{q/p}\left(\frac{v\cdot w^q - v^q\cdot w}{\beta^{2q}-\beta^2}\right).$$ 
\begin{lemma}\label{lm:commute}
Let $A$ and $B$ be elements of $P_\infty$. Then $A$ and $B$ commute if
and only if $\scal{\tau(A)}{\tau(B)}_a=0$.
\end{lemma}
\begin{proof} 
This follows from \cite{ketkar06} and the direct limit structure. 
\end{proof}

\begin{lemma}\label{lm:q2c}
Let $Q$ be an $\F_{q^2}$-linear $[(n,k,m)]_q$ quantum convolutional 
code with stabilizer $S$,
where $S=\dirlimit (S_i,\iota_{ij})$ and $S_0$ an abelian subgroup of
$P_0$ such that
\textbf{S1}, \textbf{S2}, and \textbf{S3} hold.
Then $C=\sigma^{-1}\tau(S)$ is an $\F_{q^2}$-linear
$(n,(n-k)/2;\mu\leq \ceil{m/n})_{q^2}$ convolutional code generated by
$\sigma^{-1}\tau(S_0)$. Further, $C\subseteq C^\hdual$.
\end{lemma}
\begin{proof}
Recall that $\sigma : \F_{q^2}[D]^n \rightarrow \Gamma_{q^2}$, maps $u(D)$ in
$\F_{q^2}[D]^n$ to $\sum_{i=0}^{n-1}D^iu_i(D^n)$. It is invertible, thus
 $\sigma^{-1}\tau(e) =\sigma^{-1}\circ \tau(e) $ is 
well defined for any $e$ in $P_\infty$.  
Since $S$ is generated by shifted
versions of $S_0$, it follows that $C=\sigma^{-1}\tau(S)$ is generated as the $\F_{q^2}$ span
of $\sigma^{-1}\tau(S_0)$ and its shifts, {\em i.e.}, $D^l\sigma^{-1}\tau(S_0)$, where
$l\in \N$.  
Since $Q$ is an $\F_{q^2}$-linear $[(n,k,m)]_q$ quantum convolutional code, $S_0$
defines an $[[n+m,k+m]]_q$ stabilizer code with $(n-k)/2$
$\F_{q^2}$-linear generators. Since the maps $\sigma$ and $\tau$ are
linear $\sigma^{-1}\tau(S_0)$ is also $\F_{q^2}$-linear. As
$\sigma^{-1}\tau(e)$ is in $\F_{q^2}[D]^n$ we can define an
$(n-k)/2\times n $ polynomial generator matrix that generates $C$.
This generator matrix need not be right
invertible, but we know that there exists a right invertible
polynomial generator matrix that generates this code. 
Thus $C$   is an $(n,(n-k)/2;\mu)_{q^2}$ code.
Since $S$ is abelian, Lemma~\ref{lm:commute}
and the $\F_{q^2}$-linearity of $S$ imply that $C\subseteq C^\hdual$.
Finally, observe that maximum degree of an element in $\sigma^{-1}\tau(S_0)$
is $\ceil{m/n}$ owing to  $\sigma$. Together with \cite[Lemma~14.3.8]{huffman03} 
this  implies that  the memory of $\sigma^{-1}\tau(S)$ must be
$\mu \leq \ceil{m/n}$.
\end{proof}

We define the degree of an $\F_{q^2}$-linear $[(n,k,m)]_q$ quantum
convolutional code $Q$ with stabilizer $S$ as the degree of the
classical convolutional code $\sigma^{-1}\tau(S)$. We denote an
$[(n,k,m)]_q$ quantum convolutional code with free distance $d_f$ and
total constraint length $\delta$ as $[(n,k,m;\delta,d_f)]_q$. It must
be pointed out this notation is at variance with the classical codes
in not just the order but the meaning of the parameters.

\begin{corollary}\label{co:q2c}
An $\F_{q^2}$-linear $[(n,k,m;\delta,d_f)]_q$ convolutional stabilizer
code implies the existence of an $(n,(n-k)/2;\delta)_{q^2}$
convolutional code $C$ such that $d_f=\wt(C^\hdual \setminus C)$.
\end{corollary}
\begin{proof}
As before let $C=\sigma^{-1}\tau(S)$, by Lemma~\ref{lm:commute} we can
conclude that $\sigma^{-1}\tau(C_{P_\infty}(S)) \subseteq C^\hdual$. Thus an
undetectable error is mapped to an element in $C^\hdual \setminus
C$. While $\tau$ is injective on $S$ it is not the case with
$C_{P_\infty}(S)$.  However we can see that if $c$ is in $C^\hdual
\setminus C$, then surjectivity of $\tau$ (on $C_{P_\infty}(S)$) implies that there
exists an error $e$ in $C_{P_\infty}(S)\setminus Z(P_\infty)S$ such that
$\tau(e)=\sigma(c)$. As $\tau$ and $\sigma$ are isometric $e$ is an undetectable
error with $\wt(c)$. Hence, we can conclude that $d_f=\wt(C^\hdual\setminus
C)$. Combining with Lemma~\ref{lm:q2c} we have the claim stated.
\end{proof}

An $[(n,k,m;\delta,d_f)]_q$ code is said to be a \textit{pure code} if
there are no errors of weight less than $d_f$ in the stabilizer of the
code. Corollary~\ref{co:q2c} implies that $d_f=\wt(C^\hdual\setminus C)=\wt(C^\hdual)$.

\begin{theorem}\label{th:c2qHerm}
Let $C$ be $(n,(n-k)/2,\delta;\mu)_{q^2}$ convolutional code such that $C\subseteq C^\hdual$.
Then there exists an $[(n,k,n\mu;\delta,d_f)]_q$ convolutional stabilizer code, 
where $d_f=\wt(C^\hdual\setminus C)$. The code is pure if $d_f=\wt(C^\hdual)$.
\end{theorem}
\begin{proof}[Sketch]
Let $G(D)$ be the polynomial generator matrix of $C$, with the semi-infinite generator matrix $G$ 
 defined as in equation~(\ref{eq:Gmatrix}). 
Let $C_t=\langle \sigma(G(D)),\ldots,\sigma(D^tG(D)) \rangle = \langle C_{t-1}, \sigma(D^t G(D)) \rangle$,
where $\sigma$ is applied to every row in $G(D)$.
The self-orthogonality of $C$ implies that $C_t$ is also self-orthogonal.
In particular $C_0$ defines an $[n+n\mu,(n-k)/2]_{q^2}$ self-orthogonal  
code. 
From the theory of stabilizer codes we know that there exists an abelian
subgroup $S_0\le P_0$ such that $\tau(S_0)=C_0$, 
where $P_t$ is the Pauli group over $(t+1)n+m$ qudits; in this case $m=n\mu$.
This implies that 
$\tau(I^{\otimes nt  }\otimes S_0) =\sigma(D^tG(D))$.
Define $S_t =\langle S_{t-1}, I^{\otimes nt  }\otimes S_0\rangle$, then
$\tau(S_t)=\langle\tau(S_{t-1},\sigma(D^t G(D)) \rangle$. Proceeding
recursively, we see that $\tau(S_t)= \langle \sigma(G(D)),\ldots, \sigma(D^tG(D)) \rangle=C_t$.
By Lemma~\ref{lm:commute}, the self-orthogonality of
$C_t$ implies that $S_t$ is abelian, thus \textbf{S1} holds.
Note that $\tau(S_tZ_t/Z_t)=C_t$, where $Z_t$ is the center of $P_{t}$. 
Combining this with $\F_{q^2}$-linearity of $C_t$ implies that $S_tZ_t/Z_t$ is a $(t+1)(n-k)$ 
dimensional vector space over $F_q$; hence \textbf{S2} holds. 
For \textbf{S3}, assume that $z\neq \{1\}$ is in $S_t\cap Z_t$. Then $z$ can be 
expressed as a linear combination of the generators of $S_t$. But $\tau(z)=0$
implying that the generators of $S_t$ are dependent. Thus $S_t\cap Z_t=\{ 1\}$ and 
\textbf{S3} also holds.
Thus $S=\dirlimit (S_t,\iota_{tj} )$ defines an $[(n,k,n\mu;\delta)]_q$ convolutional stabilizer code.
By definition the degree of the quantum code is the degree
of the underlying classical code. As $\sigma^{-1}\tau(S)=C$, arguing as in Corollary~\ref{co:q2c} we can show
that $\sigma^{-1}\tau(C_{P_\infty}(S))=C^\hdual$ and $d_f=\wt(C^\hdual\setminus C)$.
\end{proof}

\begin{corollary}\label{co:c2qEuclid}
Let $C$ be an $(n,(n-k)/2,\delta;\mu)_q$ code such that $C\subseteq C^\perp$. Then there
exists an $[(n,k,n\mu;\delta,d_f)]_q$ code with $d_f=\wt(C^\perp\setminus C)$. It
is pure if $\wt(C^\perp\setminus C)=\wt(C^\perp)$.
\end{corollary}
\begin{proof}
Since $C\subseteq C^\perp$, its generator matrix $G$ as in equation~(\ref{eq:Gmatrix})
satisfies $GG^T=0$. We can obtain an $\F_{q^2}$-linear
$(n,(n-k)/2,\delta;\mu)_{q^2}$ code, $C'$ from $G$ as $C'=\Gamma_{q^2}G$. 
Since $G_i\in \F_q^{(n-k)/2\times n}$ we have $GG^{\dagger}=GG^T=0$. Thus
$C'\subseteq C'^\hdual$. Further, it can checked that 
$\wt(C'^\hdual\setminus C')= \wt(C^\perp\setminus C)$.
The claim follows from Theorem~\ref{th:c2qHerm}.
\end{proof}

\begin{theorem}[Singleton bound]\label{th:qsb}
The free distance of an $[(n,k,m;\delta,d_f)]_q$ $\F_{q^2}$-linear 
pure convolutional stabilizer code is bounded by  
\begin{eqnarray}
d_f&\leq& \frac{n-k}{2}\left ( \left\lfloor \frac{2\delta}{n+k} \right\rfloor+1
\right) + \delta+1 \nonumber
\end{eqnarray}
\end{theorem}
\begin{proof}
By Corollary~\ref{co:q2c}, there exists an $(n,(n-k)/2,\delta)_{q^2}$
code $C$ such that $\wt(C^\hdual \setminus C) =d_f$, and the purity of
the code implies that $\wt(C^\hdual)=d_f$. The dual code $C^\perp$ or
$C^\hdual$ has the same degree as code \cite[Theorem~2.66]{johannesson99}.  
 Thus, $C^\hdual$ is an $(n,(n+k)/2,\delta)_{q^2}$ convolutional code with 
 free distance $d_f$.  By the generalized Singleton bound
\cite[Theorem~2.4]{smarandache01} for classical convolutional codes,
we have
\begin{eqnarray*}
d_f &\leq & \left( n - (n+k)/2 \right )\left( \left\lfloor
\frac{\delta}{(n+k)/2} \right\rfloor+1\right) + \delta + 1,
\end{eqnarray*}
which implies the claim. 
\end{proof}


\section{Convolutional RS Stabilizer Codes}\label{section:RS}

In this section we will use Piret's construction of Reed-Solomon convolutional codes
\cite{piret88} to derive quantum convolutional codes. Let
$\alpha\in \F_{q^2}$ be a primitive $n$th root of unity, where $n|q^2-1$. 
Let $w=(w_0,\ldots,w_{n-1}),\mathbf{\gamma}=(\gamma_0,\ldots,\gamma_{n-1})$ 
be in $\F_{q^2}^n$ where  $w_i\neq 0$ and  all $\gamma_i\neq 0$ are distinct.
Then the generalized Reed-Solomon (GRS) code over $\F_{q^2}^n$ 
is the code with the parity check matrix, (cf. \cite[pages~175--178]{huffman03})
\begin{eqnarray*}
 H_{\gamma,w} =\left[ \begin{array}{llll}
w_0 &w_1&\cdots &w_{n-1}\\
w_0\gamma_0 &w_1\gamma_1  &\cdots &w_{n-1}\gamma_{n-1}\\
w_0\gamma_0^2 &w_1\gamma_1^2 &\cdots &w_{n-1}\gamma_{n-1}^{2(n-1)}\\
\vdots& \vdots  &\ddots &\vdots\\
w_0\gamma_0^{t-1} &w_1\gamma_1^{2(t-1)} &\cdots &w_{n-1}\gamma_{n-1}^{(t-1)(n-1)}\\
\end{array}\right].
\end{eqnarray*}
The code is denoted by $\text{GRS}_{n-t}(\gamma,v)$, as its generator matrix is of the
form $H_{\gamma,v}$ for some $v\in \F_{q^2}^n$. It is an $[n,n-t,t+1]_{q^2}$  MDS 
code \cite[Theorem~5.3.1]{huffman03}. 
If we choose $w_i=\alpha^i$, then $w_i\neq 0$. If $\gcd(n,2)=1$, then $\alpha^2$ is also 
a primitive $n$th root of unity; thus $\gamma_i=\alpha^{2i}$ are all distinct and we have 
an $[n,n-t,t+1]_{q^2}$ GRS code with parity check matrix $H_0$, where
\begin{eqnarray*}
 H_0 =\left[ \begin{array}{ccccc}
1 &\alpha &\alpha^2 &\cdots &\alpha^{n-1}\\
1 &\alpha^3 &\alpha^6 &\cdots &\alpha^{3(n-1)}\\
\vdots& \vdots &\vdots &\ddots &\vdots\\
1 &\alpha^{2t-1} &\alpha^{2(2t-1)} &\cdots &\alpha^{(2t-1)(n-1)}
\end{array}\right].
\end{eqnarray*}
Similarly if $w_i=\alpha^{-i}$ and $\gamma_i=\alpha^{-2i}$, then we have another
$[n,n-t,t+1]_{q^2}$ GRS code with parity check matrix
\begin{eqnarray*}
 H_1 =\left[ \begin{array}{ccccc}
1 &\alpha^{-1} &\alpha^{-2} &\cdots &\alpha^{-(n-1)}\\
1 &\alpha^{-3} &\alpha^{-6} &\cdots &\alpha^{-3(n-1)}\\
\vdots& \vdots &\vdots &\ddots &\vdots\\
1 &\alpha^{-(2t-1)} &\alpha^{-2(2t-1)} &\cdots &\alpha^{-(2t-1)(n-1)}
\end{array}\right].
\end{eqnarray*}
The $[n,n-2t,2t+1]_{q^2}$  GRS code with $w_i=\alpha^{-i(2t-1)}$ and 
$\gamma_i=\alpha^{2i}$ has a parity check matrix $H^*$ that is 
equivalent to $\left[\begin{smallmatrix} H_0\\H_1\end{smallmatrix}\right]$
up to a permutation of rows.

Our goal is to show that under certain restrictions on $n$ the following semi-infinite 
coefficient matrix $H$ determines an $\F_{q^2}$-linear Hermitian self-orthogonal
convolutional code
\begin{eqnarray}\label{eq:H} 
H = \left[ \begin{array}{ccccc}
H_0 &H_1& \zero& \cdots & \cdots\\
\zero&H_0 &H_1& \zero& \cdots \\
\vdots&\vdots&\vdots &\cdots& \ddots
\end{array}\right].
\end{eqnarray}
To show that $H$ is Hermitian self-orthogonal, it is sufficient to
show that $H_0,H_1$ are both self-orthogonal and $H_0$ and $H_1$ are
orthogonal to each other. A portion of this result is contained in
\cite[Lemma~8]{grassl04}, {\em viz.}, $n=q^2-1$. 
We will prove a slightly stronger result. 

\begin{lemma}\label{lem:RS_CC_selforthogonal_np}
Let $n|q^2-1$ such that $q+1< n\leq q^2-1$ and $2 \leq \mu=2t\leq \lfloor n/(q+1)\rfloor $, then 
$$\overline{H}_0 = (\alpha^{ij})_{1\le i<\mu, 0\le j< n}\quad \text{and}
\quad \overline{H}_1 =  (\alpha^{-ij})_{1\le i<\mu, 0\le j< n} $$ 
are self-orthogonal with respect to the Hermitian inner product.
Further, $\overline{H}_0$ is orthogonal to $\overline{H}_1$.
\end{lemma}
\begin{proof}
Denote by $\overline{H}_{0,j}= (1, \alpha^j, \alpha^{2j},\cdots,
\alpha^{j(n-1)})$ and $\overline{H}_{1,j}= (1, \alpha^{-j},
\alpha^{-2j},\cdots, \alpha^{-j(n-1)})$, where $1\leq j \leq
\mu-1$. The Hermitian inner product of $\overline{H}_{0,i}$ and
$\overline{H}_{0,j}$ is given by
\begin{eqnarray*}
\langle \overline{H}_{0,i}|\overline{H}_{0,j}\rangle_h &=&\sum_{l=0}^{n-1}
\alpha^{il}\alpha^{jql} = 
\frac{\alpha^{(i+jq)n} - 1}{\alpha^{i+jq}-1},
\end{eqnarray*}
which vanishes if $i+jq\not \equiv 0 \mod n$. If $1\leq i,j \leq \mu-1 = \floor{n/(q+1)}-1$,
then $q+1 \leq i+jq\leq (q+1)\floor{n/(q+1)}-(q+1) <n $; hence, $\langle
\overline{H}_{0,i}|\overline{H}_{0,j} \rangle_h=0$. Thus,
$\overline{H}_0$ is self-orthogonal. Similarly, $\overline{H}_1$ is
also self-orthogonal. Furthermore, 
\begin{eqnarray*}
\langle \overline{H}_{0,i}|\overline{H}_{1,j}\rangle_h &=&\sum_{l=0}^{n-1}
\alpha^{il}\alpha^{-jql} = 
\frac{\alpha^{(i-jq)n} - 1}{\alpha^{i-jq}-1}.
\end{eqnarray*}
This inner product vanishes if $\alpha^{i-jq}\neq 1$ or, equivalently,
if $i-jq\not\equiv 0\mod n$. Since $1\leq i,j\leq \floor{n/(q+1)}-1\leq q-2$, we have $1\leq i\leq \floor{n/(q+1)}-1 \leq q-2$ while 
$q\leq jq\leq q\floor{n/(q+1)}-q <n$. Thus $i\not\equiv jq \mod n$ and this inner
product also vanishes, which proves the claim.
\end{proof}
Since $H_i$ is contained in $\overline{H}_i$, we obtain the following:
\begin{corollary}\label{cor:RS_CC_selforthogonal_np}
Let $2 \leq \mu=2t\leq \floor{n/(q+1)} $, where $n|q^2-1$ and $q+1<n\leq q^2-1$. Then $H_0$ and $H_1$ are
Hermitian self-orthogonal. Further, $H_0$ is orthogonal to
$H_1$ with respect to the Hermitian inner product.
\end{corollary}

Before we can construct quantum convolutional codes, we need to
compute the free distances of $C$ and $C^\hdual$, where $C$ is the code
generated by $H$.
\begin{lemma}\label{lem:CCdfree}
Let $2\leq 2t\leq \floor{n/(q+1)}$, where $\gcd(n,2)=1 $,  $n |q^2-1$ and $q+1<n\leq
q^2-1$.  Then the convolutional code $C=\Gamma_{q^2} H$ has free
distance $d_f\geq n-2t+1 >2t+1=d_f^\perp$, where 
$d_f^\perp=\wt(C^\hdual)$ is the free distance of $C^\hdual$.
\end{lemma}
\begin{proof}
Since $d_f^\perp=\wt(C^\hdual)=\wt(C^\perp)$, we compute $\wt(C^\perp)$.
Let $c=(\ldots, 0,c_0,\ldots, c_l,0,\ldots)$ be a codeword in
$C^\perp$ with $c_i\in \F_{q^2}^n$, $c_0\neq 0$, and $c_l\neq 0$.  It
follows from the parity check equations $cH^T=0$ that
$c_0H_1^T=0=c_lH_0^T$ holds.  Thus, $\wt(c_0),\wt(c_l)\geq t+1$. If
$l>0$, then $\wt(c)\geq \wt(c_0)+\wt(c_l)\geq 2t+2$.  If $l=0$, then
$c_0$ is in the dual of $H^*$, which is an $[n,n-2t,2t+1]_{q^2}$
code. Thus $\wt(c)=\wt(c_0)\geq 2t+1$ and $d_f^\perp\geq 2t+1$.  But
if $c_x$ is in the dual of $H^*$, then $(\ldots,0,c_x,0,\ldots)$ is a
codeword of $C$. Thus $d_f^\perp =2t+1$. 

Let $(\ldots,c_{i-1},c_i,c_{i+1},\ldots)$ be a nonzero codeword in
$C$. Observing the structure of $C$, we see that any nonzero $c_i$
must be in the span of $H^*$. But $H^*$ generates an
$[n,2t,n-2t+1]_{q^2}$ code. Hence $d_f\geq n-2t+1$. If $2t\leq
\floor{n/(q+1)}$, then $t\leq n/6$; thus $d_f\geq n-2t+1 >2t+1
=d_f^\perp$ holds.
\end{proof}
The preceding proof generalizes \cite[Corollary~4]{piret88} where the free distance of
$C^\perp$  was computed for $q=2^m$.

\begin{theorem}\label{th:qcc-rs}
Let $q$ be a power of a prime, $n$ an odd divisor of $q^2-1$, such that
$q+1<n\leq q^2-1$ and $2\leq \mu=2t \leq \floor{n/(q+1)}$.  
Then there exists a pure quantum convolutional code with parameters $[(n,
n-\mu, n;\mu/2,\mu+1)]_q$. This code is optimal, since
it attains the Singleton bound with equality. 
\end{theorem}
\begin{proof}
The convolutional code generated by the coefficient
matrix $H$ in equation (\ref{eq:H}) has parameters
$(n,\mu/2,\delta\leq \mu/2;1,d_f)_{q^2}$. Inspecting the corresponding
polynomial generator matrix shows that $\delta\le \mu/2$, since
$\nu_i=1$ for $1\le i\le \mu/2$. 
By Corollary~\ref{cor:RS_CC_selforthogonal_np}, this code is Hermitian
self-orthogonal; moreover, Lemma~\ref{lem:CCdfree} shows that
the distance of its dual code is given by $d_f^\perp=\mu+1 <d_f$. By
Theorem~\ref{th:c2qHerm}, we can conclude that there exists a pure
convolutional stabilizer code with parameters $[(n,n-\mu,n; \delta \le
\mu/2, \mu+1)]_q$. It follows from Theorem~\ref{th:qsb} that 
$$ 
\begin{array}{l@{\,}c@{\,}l}
\mu+1 &\le& (\mu/2)\left(\left\lfloor
2\delta/(2n-\mu)\right\rfloor+1\right)+\delta+1\\
&\le& (\mu/2)\left(\left\lfloor \mu/(2n-\mu)\right\rfloor+1\right)+\delta+1.
\end{array}
$$ Since $\left\lfloor \mu/(2n-\mu)\right\rfloor=0$, the right hand
side equals $\mu/2+\delta+1$, which implies $\delta=\mu/2$ and the optimality
of the quantum code.
\end{proof}


\section{Convolutional RM Stabilizer Codes}\label{section:RM}
In this section, we derive convolutional stabilizer codes from
quasi-cyclic subcodes of binary Reed-Muller block
codes~\cite{esmaeili97}, taking advantage of the framework developed
by Esmaeili and Gulliver for classical convolutional
codes~\cite{esmaeili98}.

Let $u=(u_1,u_2,\ldots, u_n)$ and $v=(v_1,v_2,\ldots,v_n)$ be vectors
in $\F_2^n$; we define their \textit{boolean product} as
$uv=(u_1v_1,u_2v_2,\ldots,u_nv_n).$
The product of $i$ such $n$-tuples is said to have \textit{degree}~$i$. 

Let $b_0=(1,1,\ldots,1) \in \F_2^{2^m}$. For $m>0$ and $1\leq i\leq
m$, define $b_i \in \F_2^{2^m}$ as concatenation of $2^{m-i}$ blocks
of the form $\zero\one\in \F_2^{2^i}$, where $\zero$ and $\one$ are
the constant zero and one vectors in $\F_2^{2^{i-1}}$, respectively.
Let $0\leq r<m$ and $B=\{b_1,b_2,\ldots,b_m\} \subseteq
\F_2^{2^m}$. Then the $r$th order \textit{Reed-Muller code} $\RM(r,m)$ is the
linear span of $b_0$ and all products of elements in $B$ of degree $r$
or less. The code $\RM(r,m)$ has dimension $k(r)=\sum_{i=0}^{r}\binom{m}{i}$
and minimum distance~$2^{m-r}$; the dual of $\RM(r,m)$ is given by
$\RM(r,m)^\perp=\RM(m-1-r,m)$, and the dual distance of $\RM(r,m)$ is
$2^{r+1}$, see~\cite{huffman03} for details.

Let $B_m^i$ denote the set of all products of elements in~$B$ of
degree~$i$. For $0\leq i\leq r <m$, a generator matrix $G_m^r$ of
$\RM(r,m)$ is given by (see~\cite{esmaeili98} for details)
$$
G^r_m = \left[ \begin{array}{l}B^r_m\\B^{r-1}_m\\ \quad \vdots\\ B^{i+1}_m\\
G_m^{i}\end{array} \right].
$$

Let $w_{\mu}=(110\cdots0)\in \F_2^{2^\mu}$. Let $lw_{\mu}$ denote the
vector obtained by concatenating $l$ copies of $w_{\mu}$. For $0\leq
i\leq l-1$, let $M_{i,l}= (2^{l-i-1}w_{i+1})\otimes B_{m-l}^{r-i}$,
which is a matrix of size $\binom{m-l}{r-i} \times 2^m$, and let
$M_{l,l}=\left[\begin{array}{cccc} G_{m-l}^{r-l} & \zero & \cdots&
\zero
\end{array}\right]$. 
One can derive a convolutional code as the rowspan of the
semi-infinite matrix $G$ given in~(\ref{eq:Gmatrix}), where $\mu=2^l
-1$ and the matrices $G_i$, $0\le i< 2^l$, are defined by
\begin{small}
$$\left[\begin{array}{cccccc} G_0&G_1&\cdots  & G_{2^{l}-1}
\end{array} \right]=\left[ \begin{array} {l}
M_{0,l}\\
M_{1,l}\\
\quad\vdots\\
M_{l-1,l}\\
M_{l,l}
\end{array}\right].
$$%
\end{small}%
We note that $G_0=G_{m-l}^{r}$ and that the rows of $G_i$, $1\leq
i\leq 2^l-1$, are a subset of the rows in $G_0$.  The convolutional
code generated by $G$ is a $(2^{m-l},\sum_{i=0}^r{m-l \choose i})_2$
code with free distance $2^{m-r}$, see~\cite{esmaeili98}. Even though
$G$ is corresponds to a catastrophic encoder, we can conclude the
following:
\begin{lemma}
Let $C=\Gamma_2 G$. Then $d_f^\perp$, the free distance of the
convolutional code $C^\perp$ is $2^{r+1}$.
\end{lemma}
\begin{proof}
Let $c_0$ be a codeword in the dual of $\RM(r,m-l)$ {\em i.e.},
$c_0G_0^T=0$. As $G_i$ are submatrices of $G_0$ we have $c_0G_i^T=0$.
It follows that $c=(\ldots,0,c_0,0,\ldots)$ satisfies $cG^T=0$ and is
in $C^\perp$.  Thus $d_f^\perp \leq \min \wt(c_0) =
\wt(\RM(r,m-l)^\perp)=2^{r+1}$.

Let $c=(\ldots,0,c_0,\ldots,c_l,0,\dots)$ be a codeword of minimum
weight in $C^\perp$.  Since $cG^T=0$, we can infer that
$c_0G_{2^l-1}^T=0= c_lG_{0}^T$.  Since $c_l$ is in the dual space of
$G_0$, it has a minimum weight of $2^{r+1}$.  Therefore, $\min
\{\wt(c_0)+2^{r+1}\} \leq d_f^\perp\leq 2^{r+1}$; hence
$d_f^\perp=2^{r+1}$.
\end{proof}
\begin{lemma}\label{lm:rmOrthogonality}
Let $1\leq l\leq m$ and $0\leq r\leq \lfloor (m-l-1)/2\rfloor$, then
the convolutional code generated by $G$ is self-orthogonal.
\end{lemma}
\begin{proof}
It is sufficient to show that $G_iG_j^T=0$ for $0\leq i,j\leq
2^l-1$. Since the rows of $G_i$ are a subset of the rows of $G_0$ it
suffices to show that $G_0$ is self-orthogonal. For $G_0$ to be
self-orthogonal we require that $r\leq (m-l)-r-1$ which holds. Hence,
$G$ generates a self-orthogonal convolutional code.
\end{proof}

\begin{theorem}
Let $1\leq l\leq m$ and $0\leq r\leq \lfloor (m-l-1)/2\rfloor$, then
there exist pure linear quantum convolutional codes with the
parameters $[(2^{m-l},2^{m-l}-2k(r),\le 2^{m-l}(2^{l}-1)\,)]_2$ and
free distance $2^{r+1}$, where $k(r)=\sum_{i=0}^r {m-l\choose i}$.
\end{theorem}
\begin{proof}
By Lemma~\ref{lm:rmOrthogonality}, $G$ defines a linear
self-orthogonal convolutional code with parameters $(2^{m-l},k(r),\le
2^{l}-1)_2$ and free distance $2^{m-r}$.  By
Corollary~\ref{co:c2qEuclid} there exists a linear
$[(2^{m-l},2^{m-l}-2k(r),\le 2^{m-l}(2^l-1)\,)]_2$ convolutional
stabilizer code. For $0\leq r\leq \lfloor (m-l-1)/2\rfloor$, the dual
distance $2^{r+1}< 2^{m-r}$, hence the code is pure.
\end{proof}
After circulating the first version of this manuscript, Grassl and
R\"otteler kindly pointed out that the convolutional codes in
\cite{esmaeili98} that are used here have degree 0, hence, are a
sequence of juxtaposed block codes disguised as convolutional
codes. Consequently, the codes constructed in the previous theorem
have parameters $[(2^{m-l},2^{m-l}-2k(r), 0; 0, 2^{r+1})]_2$.

\section{Conclusion}\label{sec:conclusion}
We developed an approach to convolutional stabilizer codes that is
based on a direct limit construction, formalizing the arguments given
in~\cite{ollivier04}. We proved a Singleton bound for pure
convolutional stabilizer codes, and derived an optimal family of
quantum convolutional codes attaining this bound from 
generalized Reed-Solomon codes. We illustrated how to use quasi-cyclic
subcodes of Reed-Muller codes to construct a family of convolutional
stabilizer codes; this method can be applied to other quasi-cyclic
codes as well.

\textit{Acknowledgments.} This research was supported by NSF CAREER
award CCF~0347310 and NSF grant CCF~0622201.


\end{document}